\documentclass[fleqn,usenatbib]{mnras}

\usepackage{newtxtext,newtxmath}
\usepackage[T1]{fontenc}

\DeclareRobustCommand{\VAN}[3]{#2}
\let\VANthebibliography\thebibliography
\def\thebibliography{\DeclareRobustCommand{\VAN}[3]{##3}\VANthebibliography}

\usepackage{graphicx}	% Including figure files
\usepackage{amsmath}	% Advanced maths commands
\usepackage{hyperref}
\hypersetup{
    citecolor=blue,
    colorlinks=true,
    linkcolor=blue,
    filecolor=magenta,      
    urlcolor=blue,
    pdftitle={Overleaf Example},
    pdfpagemode=FullScreen,
    }

\usepackage{xcolor}
\usepackage{textcomp}
\definecolor{dkblue}{RGB}{54, 86, 169}

\newcommand{\EQ}[1] {Eq.~(\ref{#1})}
\newcommand{\FIG}[1] {Figure~\ref{#1}}
\newcommand{\TAB}[1] {Table~\ref{#1}}

\newcommand{\SEC}[1] {Section~\ref{#1}}

\newcommand{\NO}{no.}

\def\NAME {PSR }
\title[Study of PSR J0628+0909 with FAST]{Exploring the single-pulse behaviours of PSR J0628+0909 with FAST}

\author[J.~A.~Hsu et al.]{
J.~A.~Hsu,$^{1,2,3}$
J.~C.~Jiang,$^{2,1}$\thanks{E-mail: jiangjinchen@bao.ac.cn}
H.~Xu,$^{2,4,1}$
K.~J.~Lee$^{1,2,4}$\thanks{E-mail: kjlee@pku.edu.cn}
and R.~X.~Xu$^{1,4}$\thanks{E-mail: r.x.xu@pku.edu.cn}
\\
$^{1}$Department of Astronomy, Peking University, Beijing 100871, China\\
$^{2}$National Astronomical Observatories, Chinese Academy of Sciences, 20A Datun Road, Chaoyang District, Beijing 100101, China\\
$^{3}$Jodrell Bank Centre for Astrophysics, Department of Physics and Astronomy, University of Manchester, Manchester M13 9PL, UK\\
$^{4}$Kavli Institute for Astronomy and Astrophysics, Peking University, Beijing 100871, China
}

\date{Accepted XXX. Received YYY; in original form ZZZ}

\pubyear{2022}

\begin{document}
\label{firstpage}
\pagerange{\pageref{firstpage}--\pageref{lastpage}}
\maketitle

\begin{abstract}

% The abstract should briefly describe the aims, methods, and main results of the paper.
% It should be a single paragraph not more than 250 words (200 words for Letters).
% No references should appear in the abstract.

More than 100 rotating radio transients (RRATs) have been discovered since 2006. However, it is unclear whether RRATs radiate in the nulling states. \NAME J0628+0909 has been classified as an RRAT. In this paper, we study the single pulses and integrated pulse profile of \NAME J0628+0909 to check whether we can detect pulsed radio emission in the nulling states. We also aim to study the polarisation of the RRAT and its relationship to the general pulsar population. We used the Five-hundred-meter Aperture Spherical radio Telescope (FAST) to observe \NAME J0628+0909 in the frequency range from 1.0 to 1.5 GHz. We searched for strong single pulses and looked for pulsed emission in the RRAT nulling states. Polarisation profiles, the single-pulse energy distribution, and waiting-time statistics were measured. The Faraday rotation measure and dispersion measure values are updated with the current observation. The single-pulse polarisation behaviours show great diversity, similar to the case of pulsars. Based on the integrated pulse profile and single-pulse energy statistics, we argue that continuous pulsar-like emission exists in addition to the transient-like burst emission for \NAME J0628+0909. We find that the pulse waiting-time is not correlated with the pulse energy and conclude that the strong transient emission of RRAT is not generated by the energy store-release mechanism.

\end{abstract}


%%%%%%%%%%%%%%%%%%%%%%%%%%%%%%%%%%%%%%%%%%%%%%%%%%

%%%%%%%%%%%%%%%%% BODY OF PAPER %%%%%%%%%%%%%%%%%%

\section{Introduction}

Over 100 rotating radio transients (RRATs) have been reported\footnote{see \emph{RRATalog}, a compiled list of RRATs at \url{http://astro.phys.wvu.edu/rratalog/}.} since the first discovery in 2006 \citep{McLaughlin_2006}. The definition of an RRAT is still not rigorous. \citet{SB10} defined an RRAT as \emph{``an object which emits only non-sequential single bursts with no otherwise detectable emission at the rotation period''}. Another definition was given by \citet{2011BASI...39..333K}, who referred to \emph{``a repeating radio source, with underlying periodicity, which is more significantly detectable via its single pulses than in periodicity searches''}, whilst \citet{2022arXiv220100295A} identified RRATs as \emph{``radio pulsars which can only be detected through single-pulse searches''}. In this paper, we follow the convention of \citet{2011BASI...39..333K}; that is, RRATs, a peculiar subclass of radio pulsars, manifest an extreme pulse-to-pulse variability with high \emph{nulling fractions}, namely the ratios when RRATs turn off their radio emission. Long-term RRAT monitoring enabled coherent timing solutions of RRATs to be found \citep{McLaughlin_2009, Keane_2011}, which showed that RRATs belong to a population of longer-period radio pulsars with high magnetic fields . Glitches, the sudden increase of pulsar rotation frequency, have also been detected in some RRATs \citep{2009MNRAS.400.1439L,Bhattacharyya_2018}, and the post-glitch over-recovery of the frequency derivative indicates a close relationship between RRATs and the high-magnetic-field radio pulsar population.

Two major classes of possibilities have been proposed to explain the intermittent behaviour of RRATs \citep{2009MNRAS.400.1439L}, namely (1) radio pulsar models and (2) transient X-ray magnetar models. In the first class of models, the emission of RRAT either is completely turned off in the null states \citep{Zhang_2007} or is too weak to be detected \citep{Weltevrede_2006}. In the second class of models, the radio emissions may be triggered by X-ray outbursts of usually quiescent magnetars \citep{CRH07}, a class of neutron star believed to have an extremely strong magnetic field $B\ge 10^{13}$~G where the radiation is powered by the decay of internal magnetic fields. Both classes of model are supported by observational evidence. The recent detection of continuous pulse trains from RRAT~J1913+1330 and J1538+2345 \citep{2019SCPMA..6259503L} supports the pulsar-like models, while the similar spin-down properties between RRATs and magnetars supports the transient X-ray magnetar hypothesis \citep{2008MNRAS.389.1399E, McLaughlin_2009,2009MNRAS.400.1439L}. 

Recently, MJy-level radio bursts from the magnetar SGR 1935+2154 were detected \citep{2020Natur.587...54C,2020Natur.587...59B}, and, later, the normal pulsar-like radio pulses were discovered to be approximately $10^{9}$ times weaker \citep{2020ATel13699....1Z,2020ATel14084....1Z}. The observations indicated a potential link between cosmological fast radio burst events \citep{2020Natur.587...45Z} and pulsar intermittency, which started to attract more attention. A close look at Galactic RRATs, particularly an investigation of their properties when they are in the null state, may help us to understand how FRBs channel or store energy to power the radio bursts at a power level of $10^{42}\,\mathrm{ergs\,s^{-1}}$ \citep{Luo_2018}.
   
\NAME J0628+0909 was originally discovered as single pulses in the \emph{Pulsar survey using the Arecibo L-band Feed Array} (PALFA survey; \citealt{Cordes_2006}). The pulsar has a pulse period of $P=1241.4~\mathrm{ms}$ and a dispersion measure of $\mathrm{DM}=88~\mathrm{pc~cm^{-3}}$. Follow-up ALFA observations reported a burst rate of $141~\mathrm{hr^{-1}}$ and a pulse width at half maximum of $W=10~\mathrm{ms}$ \citep{Deneva_2009}. Later, J0628+0909 was identified as an RRAT, and its precise position was measured using the Karl G. Jansky Very Large Array \citep{2012ApJ...760..124L}.

In this paper, we present an observation of \NAME J0628+0909 using the \emph{Five-hundred-meter Aperture Spherical radio Telescope} (FAST \citealt{Peng_2000, Jiang_2019}), the high sensitivity of which provides a new opportunity to characterise the pulse flux statistics of an RRAT at the low end. In \SEC{sec:obs}, we describe the setup of our observation. Data analysis, including polarimetry and statistical modelling of the single-pulse distribution, is given in \SEC{sec:result}. We note that the morphology of single pulses is very diverse, with some of them showing very high degrees of polarisation resembling that of radio magnetars \citep{2007ApJ...659L..37C,2013Natur.501..391E} or repeating FRBs \citep{Luo_2020, 2022Natur.609..685X}. Our folded data indicate a weak radiation mode for \NAME J0628+0909; that is, the pulsar can still radiate radio pulses in the `nulling' state, despite the emission being three orders of magnitude weaker than the transient pulse emission. We provide a discussion and present conclusions in \SEC{sec:discuss}.

\section{Observations and data processing}\label{sec:obs}
We observed \NAME J0628+0909 with FAST, for which the effective gain after correcting the aperture efficiency is $G\simeq 16\,\mathrm{K\,Jy^{-1}}$. Our observation was performed with the L-band 19-beam receiver installed at the main focal point \citep{Jiang_2020}, which covers the frequency range of 1.0 to 1.5 GHz with a typical system noise temperature of $T_{\mathrm{sys}} \approx 20\,\mathrm{K}$. We recorded the 1.0 -- 1.5 GHz data with a \textsc{roach2} board-based digital backend \citep{parsons2006petaop}, where radio-frequency (RF) data were sampled in 8-bit format. Channelised filterbank data were formed in the \textsc{roach2} using Field-Programmable Gate Array (FPGA)-based polyphase filterbanks and then transferred to the data recording computer cluster. The data were stored in 8-bit format at the rate of $49.152~\mathrm{\mu s}$ per sample with frequency resolutions of 0.122 MHz (i.e. 4096 channels for 500 MHz bandwidth). The observation started on 2021 June 4 07:06:51 UTC and lasted for 29 minutes. Before the observation, we also observed 1 min of the polarisation calibration signal from noise diode injection.

\begin{figure}
\centering
\includegraphics[width=\hsize]{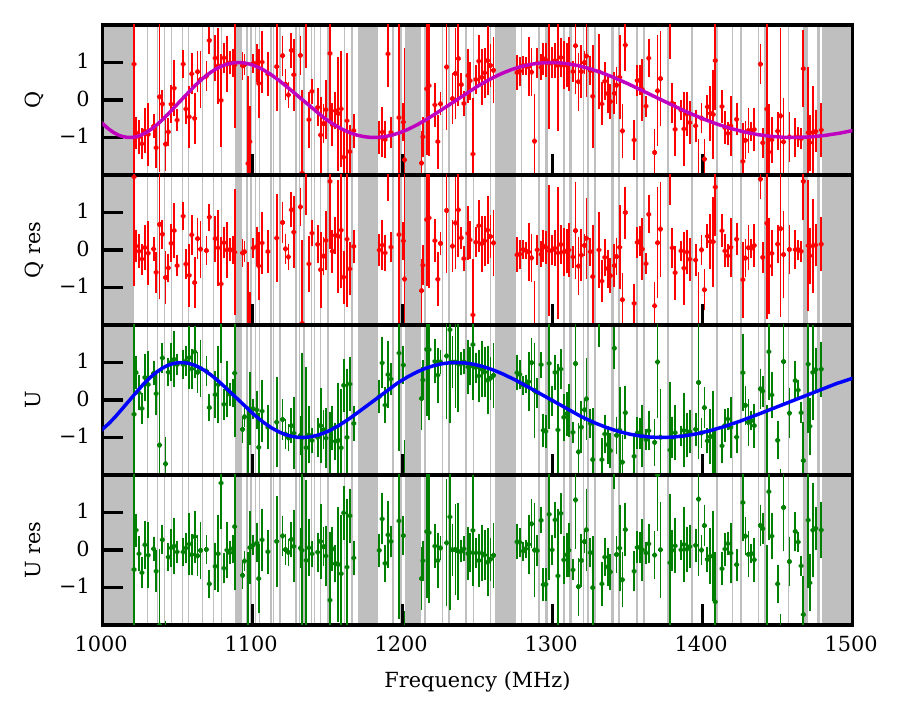}
  \caption{$Q-U$ fitting for the rotation measure. The $x$-axis is the radio frequency, and the $y$-axes are normalised Stokes $Q$, $U$ and the fitting residuals as labelled. The dots with 68\% confidence-level error bars are the observed values, and solid curves are from the best-fitting model. Grey shaded areas indicate the channels removed in the radio-frequency interference mitigation stage.
  \label{fig:Q-U}}
\end{figure}

We refined the DM by fitting the time-of-arrivals (TOAs) to the cold plasma dispersion relation using \textsc{tempo2} \citep{2006MNRAS.369..655H}, where TOAs were generated for eight subbands spread equally across the frequency range of 1.0 -- 1.5 GHz. We obtained $\mathrm{DM}=(88.47\pm 0.06)~\mathrm{pc~cm^{-3}}$. The data were then  de-dispersed at the refined DM value. Our DM measurement is fully consistent with the previously published value of $\mathrm{DM}=88.3~\mathrm{pc~cm^{-3}}$ \citep{Nice_2013}. After de-dispersion, we performed polarisation calibration using the noise diode signal, which was injected to mimic the 45-degree linearly polarised white noise. The polarisation calibration was performed with the single-axis model \citep{PSRCHIVE}, which provided an accuracy of $\sim0.5$\% according to the lab-measured specification of the feed \citep{Dunning_2017}. We also corrected the Faraday rotation effect with the rotation measure ($\rm RM$) derived by performing the $Q-U$ fitting technique \citep{Desvignes_2019,Luo_2020} on the time-integrated data of periods with a single-pulse signal-to-noise ratio $\mathrm{S/N}>7$. The best-fitting value is $\mathrm{RM}=140.9\substack{+1.5 \\ -1.4}~\mathrm{rad~m^{-2}}$ after implementing the ionospheric correction computed with the software package \textsc{ionFR} \citep{ionfr}. The $Q-U$ fitting and residuals are shown in \FIG{fig:Q-U}.

We folded our data with the timing ephemeris provided by \citet{Nice_2013}, where the ephemeris parameters with an updated DM value are reproduced in ~\TAB{tab:ephemerides}. The software package \textsc{dspsr} \citep{vanStraten_2011} was used in the folding process. As the timing ephemeris is 10 years old, we validated the ephemerides by checking the phase drift of pulses during our half-hour observation. As no significant phase drift was found, we believe that the ephemeris is accurate enough for the current study.

\begin{table}
\caption{\label{tab:ephemerides}
Ephemeris for \NAME J0628+0909. All the values and errors are from \citet{Nice_2013}, except for DM and RM, which were measured using data in this paper.}
\begin{tabular}{ll}
\hline\hline
Right Ascension (RA, J2000)            & 06:28:36.183(5)   \\
Declination (Dec, J2000)           & +09:09:13.9(3)  \\
Reference epoch (PEPOCH, MJD) & 54990 \\
Rotation frequency ($F_0$)      & 0.8055282493188(20) $\rm Hz$ \\
Frequency derivative ($F_1$) & $-3.5552(13)\times10^{-16}$ $\rm s^{-2}$ \\
DM                    & $88.47\pm0.06$ $\rm cm^{-3}\, pc$ \\
RM                    & $140.9^{+1.5}_{-1.4}$ $\rm rad\, m^{-2}$ \\
\hline\hline
\end{tabular}
\end{table}

We visually inspected the de-dispersed dynamic spectra and folded subintegrations. We removed channels and subintegrations contaminated by the radio-frequency interference (RFI). We also checked to ensure that the individual pulses showed the correct dispersion signatures. To avoid the possible spectral mirror effect due to the spectral leakage \citep{harris2021multirate}, 20-MHz band edges on both sides of the bandpass were removed.

\section{Results}\label{sec:result}
\subsection{Integrated profile}
\begin{figure}
\centering
\includegraphics[width=0.99\hsize]{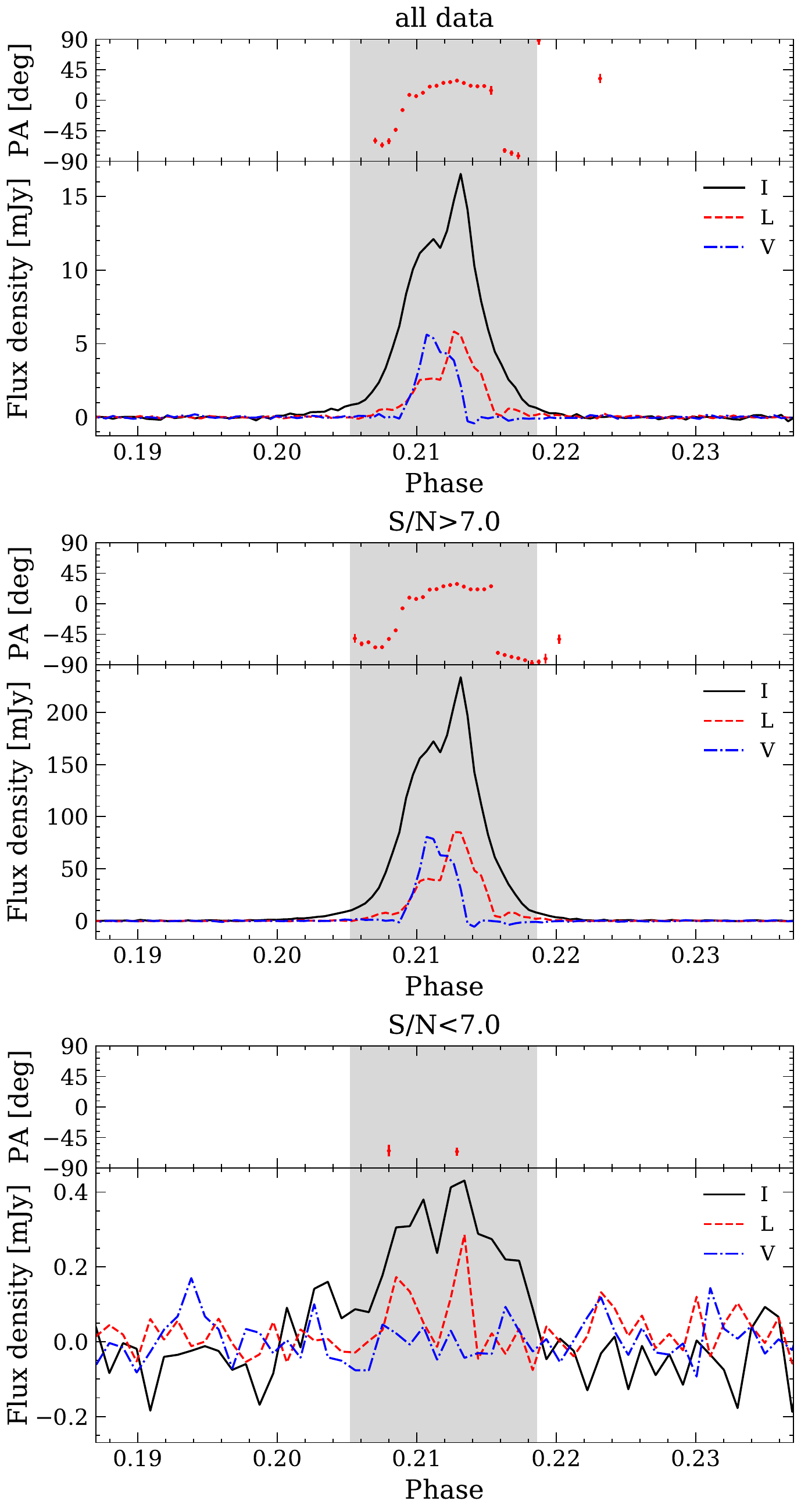}
  \caption{Integrated profiles and PA for all data (top panel, 28.4-min integration), for all single pulses with $\mathrm{S/N}\ge7$ (middle panel, 2.4-min integration), and for all data removing single pulses with $\mathrm{S/N}\ge7$ (bottom panel, 26-min integration). The top and bottom parts of each panel show the swing of PA and the polarimetric profile as a function of the pulse phase. The total intensity ($I$) is represented by black solid curves, while the linear polarisation ($L$) and circular polarisation ($V$) are represented by red dashed and blue dash-dotted curves, respectively. The PA swing shows an abrupt orthogonal jump at the pulse phase 0.215. The grey shaded area is the on-pulse region which is defined as twice the width producing the best $\mathrm{S/N}$.
  \label{fig:inte_profile}}
\end{figure}

The integrated polarisation pulse profile (over both frequency and time) are shown in \FIG{fig:inte_profile}, where three different kinds of pulse profiles are given: (1) the integrated pulse profile of all data, (2) the integrated pulse profiles of \emph{only} the individual pulses with $\mathrm{S/N}\ge 7$, and (3) the integrated pulse profile of all data excluding the individual pulses with $\mathrm{S/N}\ge 7$, namely the integrated pulse profile of all individual pulses with $\mathrm{S/N}<7$. Our $\mathrm{S/N}$ is defined using a boxcar-matched filter \citep{MLC19}. However, because the pulse phase of \NAME J0628+0909 is confined nearly in the same phase as the integrated pulse phase (see below in section~\ref{sec:sigps}), we fixed the phase range and the width of the boxcar filter when computing the subintegration $\mathrm{S/N}$, which is defined as 
\begin{equation}
    \mathrm{S/N}=\frac{\sum_{\rm{box}}{A_i}}{\sqrt{w}\cdot\sigma}\,.
    \label{eq:SN}
\end{equation}
Here, the pulse width is $w$ (in the unit of the number of bins), $\sigma$ is the off-pulse root-mean-square (rms) noise level, and the summation for pulse flux ($A_i$) is over the phase range defined by the pulse width. 

The optimal parameters of the phase range and $w$ were found by searching over the parameter space to maximise the $\mathrm{S/N}$ for the \emph{integrated pulse profile using all data}. The width ($w$) producing the best $\mathrm{S/N}$ is 8 ms or 0.7\% in phase, as indicated in \FIG{fig:inte_profile}. We also measured the pulse widths at 50\% and 10\% of the pulse profile peak, namely $W_{50}$ and $W_{10}$, as 6.7 and 12.9 ms, respectively. These values roughly agree with previous work \citep{Posselt_2021}, which round $W_{50}=9.7\pm0.6~\rm ms$ centred at 1.27 GHz with a bandwidth of 775 MHz.

We estimate the pulsar phase-average flux using the radiometer equation;
\begin{equation}
    S_{\rm mean}=\dfrac{T_{\rm sys} \mathrm{S/N}}{G\sqrt{n_{\rm pol}{t_{\rm obs}}\Delta \nu}}\sqrt{\dfrac{\delta}{1-\delta}}\,,
\label{eq:radiometer}
\end{equation}
where $T_{\rm sys}\simeq 20\,{\rm K}$ is the system noise temperature, $G\simeq 16\,\mathrm{K\, Jy^{-1}}$ is the gain of the telescope, $n_{\rm pol}=2$ represents dual-polarisation data, $t_{\rm obs}$ is the observation time, $\delta$ is the duty cycle, and $\Delta\nu$ is the observed bandwidth of 362 MHz after excluding frequency channels contaminated by RFIs. The mean flux density over the whole spin phase is therefore $54\pm 10\,\mathrm{\mu Jy}$ considering a 20\% systematic error \citep{Jiang_2019}. \citet{Nice_2013} reported a similar value: the mean flux density at 1.4 GHz $S_\mathrm{1400}=58(3)\,\mathrm{\mu Jy}$ after fitting four subbands' flux densities to a power law model.

We fitted the integrated PA curves of all data and periods with a single-pulse $\mathrm{S/N}\ge7$ to the rotating vector model (RVM; \citealt{1969Natur.221..443R,1970Natur.225..612K}) using the Bayesian method \citep{Desvignes_2019}. The best-fitting parameters are in Table~\ref{tab:rvm}. However, owing to the limited phase range, we found that the fitting does not lead to reasonable constraints for geometrical parameters; that is, the inclination angle and impact angle are strongly correlated in the posterior distributions.
\begin{table}
    \centering
    \caption{Rotating vector model fitting results for the PA curves. $\alpha$ is the inclination angle between the magnetic pole and the spin axis. $\zeta$ is the viewing angle between the line of sight and the spin axis. $\phi_0$ is the phase offset, and $\Psi_0$ is the PA offset.}
    \label{tab:rvm}
    \begin{tabular}{c c c c c}
        \hline
        \hline
        Selection & $\alpha$ ($^\circ$) & $\zeta$ ($^\circ$) & $\phi_0$ & $\Psi_0$ ($^\circ$) \\
        \hline
        All & $54\substack{+11\\-7}$ & $126\substack{+7\\-11}$ & $-0.2922\substack{+0.0003\\-0.0004}$ & $-46.8\substack{+4.2\\-3.1}$\\
        $\mathrm{S/N}\ge7$ & $83\substack{+4\\-9}$ & $97\substack{+9\\-4}$ & $-0.29243\substack{+0.00015\\-0.00018}$ & $-45.2\substack{+1.1\\-0.7}$\\
        \hline
        \hline
    \end{tabular}
    \label{tab:my_label}
\end{table}

\subsection{Single-pulse properties}
\label{sec:sigps}
\begin{figure}
\centering
\includegraphics[width=\hsize]{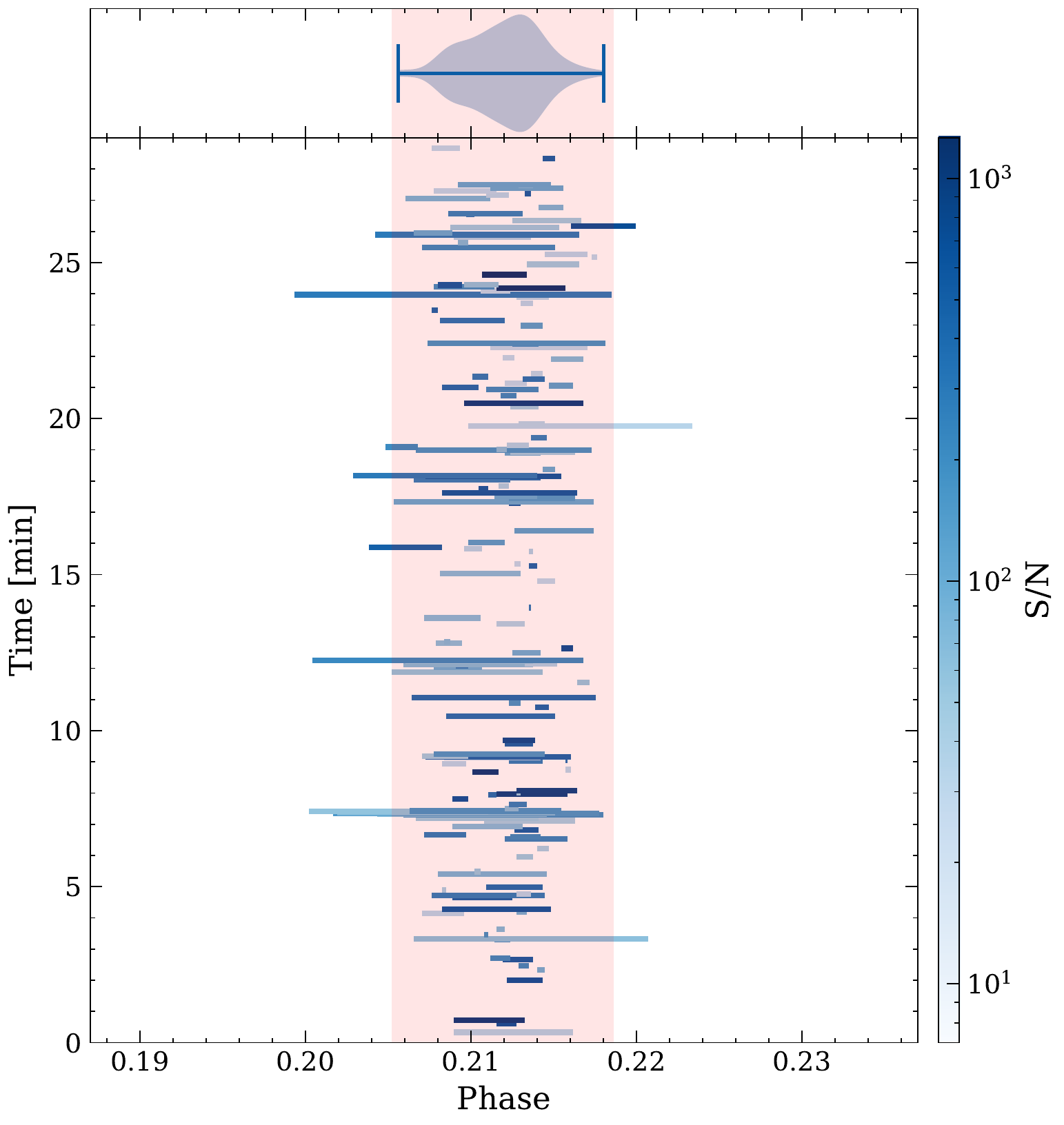}
  \caption{The time and phase distribution of all single pulses with $\mathrm{S/N}\ge7$. The $x$-axis is the pulse phase. Top panel: the distribution of the central phase of single pulses as a function of the pulse phase. Bottom panel: The $y$-axis is the time. Each horizontal blue bar represents a single pulse and its phase coverage. The colour of each horizontal bar indicates the $\mathrm{S/N}$, and the corresponding values can be read off from the colour bar on the right-hand side. Except for a few wide pulses, most of the pulses fall in the on-pulse region. The pink shaded area is the on-pulse region also shown in \FIG{fig:inte_profile}. 
  \label{fig:violinplot}}
\end{figure}

This paper uses two single-pulse search schemes to characterise the single-pulse population. We first carried out a blind single-pulse search, where the moving boxcar filter was applied to the de-dispersed time series, and the boxcar widths and centres were allowed to be the \emph{free} parameters in searching. The threshold for detecting a single pulse is $\mathrm{S/N}\ge7$ \citep{ZXM21} to reduce coloured noise artefacts. To further mitigate the possible RFI contamination, a visual inspection was performed on all single pulses, where we verified that they all had the expected dispersion signatures. In total, 155 single pulses were detected. \FIG{fig:violinplot} shows the distribution over time and the phases of all the detected single pulses. As can be seen, the single-pulse phases are relatively stable for the RRAT population, consistent with the prediction \citep{Weltevrede_2006} and later observations \citep{SB10,CBM17}. On the other hand, the widths of single pulses vary significantly, from 0.3 to 20 ms.

\begin{figure*}
\centering
\includegraphics[width=\textwidth]{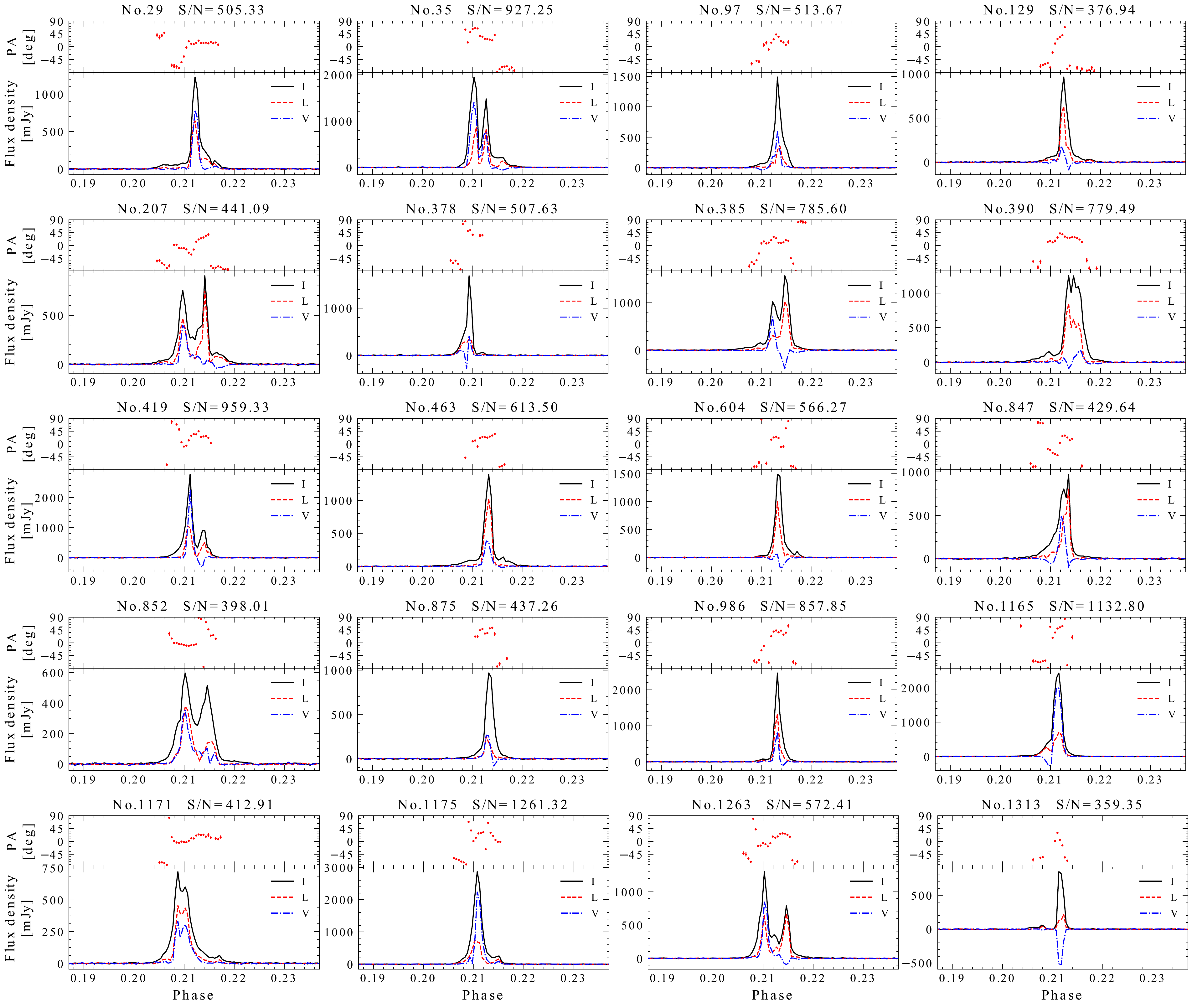}
  \caption{Sample of single pulses with the top-20 highest $\mathrm{S/N}$. For each panel, the top and bottom parts are for the PA curve and polarisation pulse profile, where, similar to in \FIG{fig:inte_profile} the solid black curve is for the total intensity, and the blue and red ones are for circular and linear polarisations, respectively. The pulse number and $\mathrm{S/N}$ are labelled at the top of each panel. Dispersion and Faraday rotation are all corrected here.
  \label{fig:SN_max20}}
\end{figure*}

The polarisation pulse profiles and dynamic spectra of the top-20 highest $\mathrm{S/N}$ single pulses are shown in \FIG{fig:SN_max20}. The single-pulse polarisation profiles are notably different from the integrated profile. Although the single-pulse PA curves more-or-less resemble that of the integrated pulse profile, there can be great diversity in the profile widths, structure morphologies, and polarisation properties. Examples include: narrow pulse (\NO 986) vs wide pulse (\NO 852); single-peak (\NO 29) vs double-peak (\NO 35) profiles; PA swing (\NO 1165) vs nearly constant PA (\NO 390), high fractional linear polarisation (\NO 604) vs high circular polarisation (\NO 1175), with and without circular polarisation sign change  (\NO 419 vs 463). Similar features have been noted in millisecond pulsars \citep{PPM22}.

\begin{figure}
\centering
\includegraphics[width=\hsize]{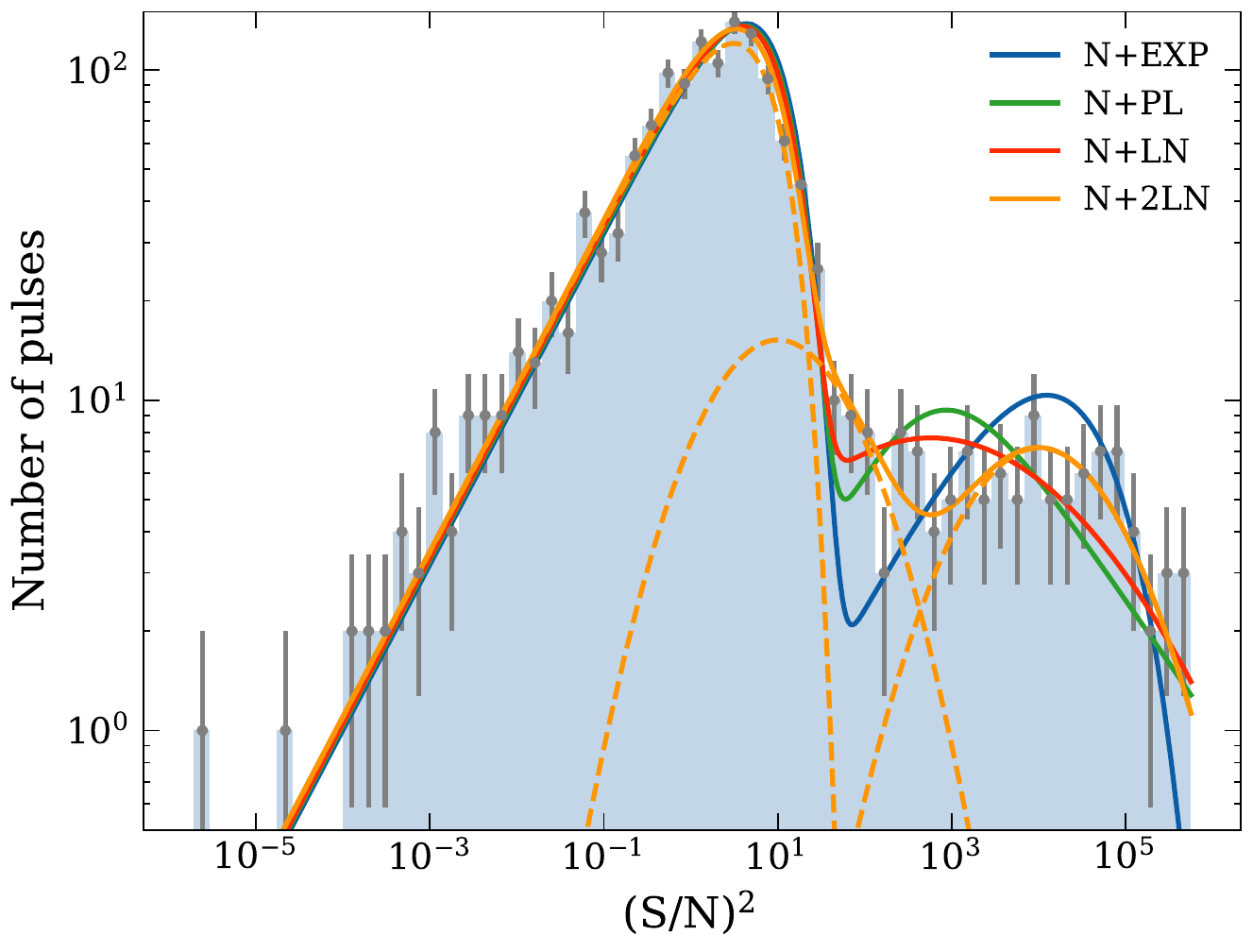}
  \caption{Distribution of $\left(\mathrm{S/N}\right)^2$ on the bilogarithmic scale. The best fit using a normal plus two log-normals (N + 2LN) is shown by the orange curve, whereas the blue, green, and red curves are for N + EXP, N + PL, and N + LN, respectively. The error bars on each bin are Poissonian uncertainties. Note that our modelling is for $\mathrm{S/N}$, but we show the distribution in the representation of $\ln\left[\left(\mathrm{S/N}\right)^2\right]$ for better visualisation of the low-$\mathrm{S/N}$ population. The dashed curves show individual components of the N + 2LN distribution.
  \label{fig:SN-fitting}}
  
\end{figure}

The blind search can be affected by random noise in the low-$\mathrm{S/N}$ regime \citep{ZXM21}. To fully characterise the single-pulse behaviour in the low-$\mathrm{S/N}$ regime, the second single-pulse detection scheme was used, where we fixed the boxcar width $w$ and centre according to the on-pulse region and computed the $\mathrm{S/N}$ without searching. The $\mathrm{S/N}$ provides a statistical description for the single-pulse strength and noise properties. For a better visual representation, we plot the histogram of $(\mathrm{S/N})^2$ on a logarithmic scale in \FIG{fig:SN-fitting}. Here, the square is introduced, as $\mathrm{S/N}$ may be negative in this case. 

We modelled the $\mathrm{S/N}$ distribution with four models and then evaluated the ''goodness of fit''. The models are all mixture probability models, and the distributions are composed of more than one distribution function. The models are mixtures of: (1) Gaussian normal and exponential distributions (N + EXP), (2) Gaussian normal and power law distributions (N + PL); (3) Gaussian normal and log-normal distributions (N + LN); and (4) Gaussian normal and two log-normal distributions (N + 2LN). The probability distribution functions are
\begin{equation}
    f_{\rm N+EXP}\left(x\right) = sf_{\rm N}\left(x\mid\mu,\sigma\right) + \left(1-s\right)f_{\rm EXP}\left(x\mid\lambda\right),
\end{equation}
\begin{equation}
    f_{\rm N+PL}\left(x\right)=sf_{\rm N}\left(x\mid\mu,\sigma\right)+\left(1-s\right)f_{\rm PL}\left(x\mid x_0, \alpha\right),
\end{equation}
\begin{equation}
    f_{\rm N+LN}\left(x\right)=sf_{\rm N}\left(x\mid\mu_1,\sigma_1\right)+\left(1-s\right)f_{\rm LN}\left(x\mid\mu_2,\sigma_2\right),
\end{equation}
\begin{equation}
\begin{aligned}
f_{\mathrm{N}+2 \mathrm{LN}}\left(x\right) &=s_{1} s_{2} f_{\mathrm{N}}\left(x \mid \mu_1, \sigma_1\right)+s_{1}\left(1-s_{2}\right) f_{\mathrm{LN}}\left(x \mid \mu_{2}, \sigma_{2}\right) \\
&+\left(1-s_{1}\right) f_{\mathrm{LN}}\left(x \mid \mu_{3}, \sigma_{3}\right),
\end{aligned}
\end{equation}
where  $s$, $s_1$, and $s_2$ are the mixture weights, $f_{\rm N}\left(x\mid\mu,\sigma\right)$, $f_{\rm EXP}\left(x\mid\lambda\right)$, $f_{\rm PL}\left(x\mid x_0, \alpha\right)$ and $f_{\rm LN}\left(x\mid\mu,\sigma\right)$ are the probability density distribution functions of Gaussian normal, exponential, power law and log-normal distributions, respectively, defined as
\begin{equation}
    \displaystyle {f_{\rm N}\left(x\mid\mu,\sigma\right)={\frac {1}{{\sqrt {2\pi }}\sigma}}\exp\left[-{\frac{\left(x-\mu\right)^{2} }{2\sigma^2}}\right]},
    \label{eq:N}
\end{equation}
\begin{equation}
    \displaystyle {f_{\rm EXP}\left(x\mid\lambda\right)= \begin{cases}\lambda\exp\left[-\lambda x\right] & x \geq 0, \\ 0 & x<0,\end{cases}}
    \label{eq:EXP}
\end{equation}
\begin{equation}
    \displaystyle {f_{\rm PL}\left(x\mid x_0,\alpha\right)= \begin{cases}\frac{C}{\left(x^2+x_0^2\right)^{\alpha/2}} & x>0, \\ 0 & x\leq0,\end{cases}}
\end{equation}
\begin{equation}
    \displaystyle {f_{\rm LN}\left(x\mid\mu,\sigma\right)= \begin{cases}\frac{1}{\sqrt{2\pi}x \sigma} \exp \left[-\frac{(\ln x-\mu)^{2}}{2 \sigma^{2}}\right] & x \geq 0, \\ 0 & x<0,\end{cases}}
\label{eq:LN}
\end{equation}
in which $C=2 \pi^{-1/2} x_0^{\alpha-1} \Gamma\left(\frac{\alpha}{2}\right)/\Gamma\left(\frac{\alpha}{2}-\frac{1}{2}\right) $ is the normalisation factor, and a corner cut-off $x_0$ is introduced to regularise the power law distribution at the low end. For each model, we used the standard Bayesian method \citep{sivia2006data} to infer the model parameters, and the software package \textsc{multinest} \citep{Feroz_2009} was used to perform the posterior sampling and to compute the Bayes factors. We understand that it is mathematically not well defined to compare the Bayes factors here directly, because the four models are \emph{not} nested \citep{casella2009consistency}. Thus, the Kolmogorov-Smirnov (KS) test was also used to check the compatibility of probability models. The maximal likelihood estimator of parameters, Bayes factors, and the KS test $p$-values are listed in \TAB{tab:SN-fitting}. We find that the N + 2LN model describes the $\mathrm{S/N}$ distribution \emph{substantially} better than the other three models, although the KS test shows that all four models are at the acceptable level given the data set. The models are compared with the measured $\mathrm{S/N}$ distribution in \FIG{fig:SN-fitting}.

\begin{table*}
\centering
\caption{\label{tab:SN-fitting}
Inferred model parameters for the $\mathrm{S/N}$ distribution.}
\begin{tabular}{cccccccccc}
\hline\hline \\
\multicolumn{10}{c}{N+EXP ($\rm M_1$)}  \\ [1.5ex]
$\mu$ & $\sigma$& \multicolumn{4}{c}{$\lambda$} & \multicolumn{2}{c}{$s$} & $p$-value & $\log_{10}{\cal B}$ \\ [1.5ex]
\hline \\
$0.31\substack{+0.12\\-0.11}$ & $1.92\substack{+0.09\\-0.09}$ & \multicolumn{4}{c}{$0.009\substack{+0.002\\-0.002}$} & \multicolumn{2}{c}{$0.91\substack{+0.02\\-0.02}$} & 0.46 & -9.02 ($\rm M_1$/$\rm M_4$) \\ [1.5ex]
\hline\hline \\
\multicolumn{10}{c}{N+PL ($\rm M_2$)} \\ [1.5ex]
$\mu$ & $\sigma$ & \multicolumn{2}{c}{$x_0$} & \multicolumn{2}{c}{$\alpha$}& \multicolumn{2}{c}{$s$} & $p$-value & $\log_{10}{\cal B}$ \\ [1.5ex]
\hline \\
$0.27\substack{+0.13\\-0.15}$ & $1.86\substack{+0.10\\-0.11}$ & \multicolumn{2}{c}{$25.83\substack{+24.15\\-15.83}$}& \multicolumn{2}{c}{$1.77\substack{+0.44\\-0.26}$} & \multicolumn{2}{c}{$0.89\substack{+0.02\\-0.03}$} & 0.58 & -6.88 ($\rm M_2$/$\rm M_4$) \\ [1.5ex]
\hline\hline \\
\multicolumn{10}{c}{N+LN ($\rm M_3$)} \\ [1.5ex]
$\mu_1$ & $\sigma_1$& \multicolumn{2}{c}{$\mu_2$} & \multicolumn{2}{c}{$\sigma_2$}& \multicolumn{2}{c}{$s$} & $p$-value & $\log_{10}{\cal B}$ \\ [1.5ex]
\hline \\
$0.23\substack{+0.14\\-0.17}$ & $1.80\substack{+0.11\\-0.10}$ & \multicolumn{2}{c}{$3.18\substack{+0.63\\-1.20}$} & \multicolumn{2}{c}{$1.87\substack{+0.69\\-0.39}$} & \multicolumn{2}{c}{$0.88\substack{+0.03\\-0.07}$} & 0.89 & -1.63 ($\rm M_3$/$\rm M_4$) \\ [1.5ex]
\hline\hline \\
\multicolumn{10}{c}{N+2LN ($\rm M_4$)} \\ [1.5ex]
$\mu_1$ & $\sigma_1$& $\mu_2$ & $\sigma_2$& $\mu_3$ & $\sigma_3$& $s_1$ & $s_2$ & $p$-value & $\log_{10}{\cal B}$ \\ [1.5ex]
\hline \\
$0.06\substack{+0.28\\-0.42}$ & $1.74\substack{+0.18\\-0.23}$ & $1.16\substack{+1.81\\-0.53}$ & $0.97\substack{+1.39\\-0.72}$ & $4.61\substack{+1.10\\-0.96}$ & $1.04\substack{+0.65\\-0.83}$ & $0.94\substack{+0.05\\-0.04}$ & $0.87\substack{+0.11\\-0.15}$ & 0.92 & 0 ($\rm M_4$/$\rm M_4$) \\ [1.5ex]
\hline\hline
\end{tabular}
\end{table*}

\begin{figure}
\centering
\includegraphics[width=\hsize]{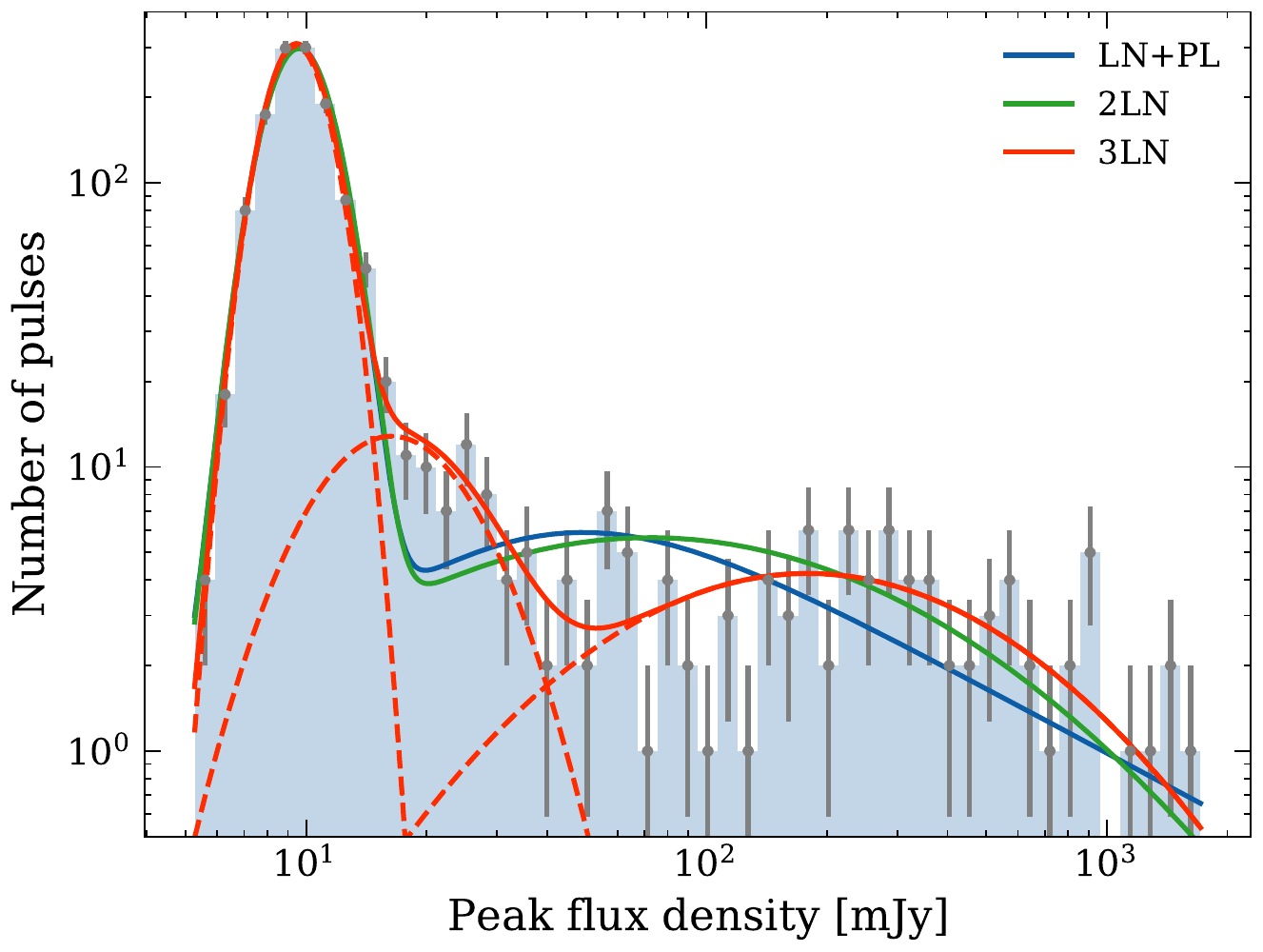}
  \caption{Peak flux density distribution on the bilogarithmic scale. The log-normal plus a power law (LN+PL, blue) and mixture models of two log-normal functions (2LN, green), and three log-normal functions (3LN, red) are shown. The error bars on each bin are Poissonian. Similar to in \FIG{fig:SN-fitting}, we convert the distribution functions to the logarithmic scale. The dashed curves show individual components of the 3LN distribution.
  \label{fig:flux-fitting}}
\end{figure}

We can compute the single-pulse peak flux density $S_{\rm peak}$ using the radiometer equation
\begin{equation}
    S_{\rm peak}=\dfrac{T_{\rm sys} {\mathrm{S/N}_{\rm peak}}}{G\sqrt{n_{\rm pol}\tau\Delta\nu}},
\end{equation}
in which $T_{\rm sys}$, $G$, $n_{\rm pol}$ and $\Delta\nu$ are the same as in \EQ{eq:radiometer}, whereas $\tau$ is the sample time. 
%The approach to calculate 
The peak signal-to-noise ratio ($\mathrm{S/N_{peak}}$) is defined as
\begin{equation}
    \mathrm{S/N}_{\rm peak}=\dfrac{A_{\rm peak}}{\sigma},
\end{equation}
where $\sigma$ is the rms value of the off-pulse profile amplitude, and $A_{\rm peak}$ is the maximum within the pulse window \citep{CBM17}. The peak flux density distribution is drawn in \FIG{fig:flux-fitting}. We modelled the peak flux distribution with three mixture models, namely the mixture models of log-normal and power law (LN + PL), and two and three log-normal functions (2LN and 3LN). The models are
\begin{equation}
    f_{\rm{LN+PL}}\left(x\right)=sf_{\rm{LN}}\left(x\mid\mu,\sigma\right)+\left(1-s\right)f_{\rm{PL}}\left(x\mid x_0,\alpha\right),
\end{equation}
\begin{equation}
\begin{aligned}
    f_{\rm{2LN}}\left(x\right) = sf_{\rm{LN}}\left(x\mid\mu_1,\sigma_1\right) + \left(1-s\right)f_{\rm{LN}}\left(x\mid\mu_2,\sigma_2\right),
\end{aligned}
\end{equation}
\begin{equation}
\begin{aligned}
    f_{\rm{3LN}}\left(x\right) & = s_1s_2f_{\rm{LN}}\left(x\mid\mu_1,\sigma_1\right) + s_1\left(1-s_2\right)f_{\rm{LN}}\left(x\mid\mu_2,\sigma_2\right) \\
    & + \left(1-s_1\right)f_{\rm{LN}}\left(x\mid\mu_3,\sigma_3\right).
\end{aligned}
\end{equation}
We performed parameter inference similar to the case for $\mathrm{S/N}$, and the inferred parameters are given in \TAB{tab:flux-fitting}. Bayes factor values indicate that we should use the mixture of three log-normal functions (3LN) to describe the distribution, although, again, the KS test indicates that all three models are acceptable for the current data set.

Scintillation caused by the interstellar medium may affect the observed intensity. The Galactic electron-density model \textsc{NE2001} \citep{2002astro.ph..7156C} estimates the scintillation bandwidth to be 67 kHz at 1 GHz, and the scintillation time is 105 s at 1 GHz for \NAME J0628+0909. As our channel bandwidth is larger than the scintillation bandwidth, and the full bandwidth is three more orders of magnitude higher, we expect that the measured single pulse fluxes will not be affected by the diffractive scintillation. On the other hand, although the total observation length is 30 times longer than the scintillation time, the observation length is still two to three orders of magnitude shorter than the refractive scintillation time-scale, which may lead to a bias in the measured single-pulse energy distribution owing to the current single-epoch observation.

\begin{table*}
\centering
\caption{\label{tab:flux-fitting}
Inferred model parameters for the distribution of peak flux density.}
\begin{tabular}{cccccccccc}
\hline\hline \\
\multicolumn{10}{c}{LN+PL ($\rm M_1$)} \\ [1.5ex]
$\mu$ [mJy] & $\sigma$ [mJy] & \multicolumn{2}{c}{$x_0$ [mJy]} & \multicolumn{2}{c}{$\alpha$} & \multicolumn{2}{c}{$s$}  & $p$-value & $\log_{10}{\cal B}$ \\ [1.5ex]
\hline \\
$2.26\substack{+0.01\\-0.01}$ & $0.19\substack{+0.01\\-0.01}$ & \multicolumn{2}{c}{$42.86\substack{+34.23\\-15.81}$} & \multicolumn{2}{c}{$1.76\substack{+0.31\\-0.16}$} & \multicolumn{2}{c}{$0.87\substack{+0.02\\-0.03}$} & $0.35$   & $-15.21$ ($\rm M_1$/$\rm M_3$) \\ [1.5ex]
\hline\hline \\
\multicolumn{10}{c}{2LN ($\rm M_2$)} \\ [1.5ex]
$\mu_1$ [mJy] & $\sigma_1$ [mJy] & \multicolumn{2}{c}{$\mu_2$ [mJy]} & \multicolumn{2}{c}{$\sigma_2$ [mJy]} & \multicolumn{2}{c}{$s$} & $p$-value & $\log_{10}{\cal B}$ \\ [1.5ex]
\hline \\
$2.26\substack{+0.01\\-0.02}$ & $0.19\substack{+0.01\\-0.01}$ & \multicolumn{2}{c}{$4.28\substack{+0.28\\-0.31}$} & \multicolumn{2}{c}{$1.42\substack{+0.21\\-0.16}$} & \multicolumn{2}{c}{$0.87\substack{+0.02\\-0.03}$} & $0.27$ & $-12.51$ ($\rm M_2$/$\rm M_3$) \\ [1.5ex]
\hline\hline \\
\multicolumn{10}{c}{3LN ($\rm M_3$)} \\ [1.5ex]
$\mu_1$ [mJy] & $\sigma_1$ [mJy] & $\mu_2$ [mJy] & $\sigma_2$ [mJy] & $\mu_3$ [mJy] & $\sigma_3$ [mJy] & $s_1$ & $s_2$ & $p$-value & $\log_{10}{\cal B}$ \\ [1.5ex]
\hline \\
$2.25\substack{+0.02\\-0.01}$ & $0.18\substack{+0.02\\-0.01}$ & $2.79\substack{+0.43\\-0.25}$ & $0.44\substack{+0.28\\-0.16}$ & $5.19\substack{+0.66\\-0.53}$ & $1.11\substack{+0.36\\-0.42}$ & $0.93\substack{+0.03\\-0.03}$ & $0.90\substack{+0.05\\-0.06}$ & $0.96$   & $0$ ($\rm M_3$/$\rm M_3$)  \\ [1.5ex]
\hline\hline
\end{tabular}
\end{table*}

\subsection{Weak-pulse analysis}\label{sec:weak}
\begin{figure}
\centering
\includegraphics[width=\hsize]{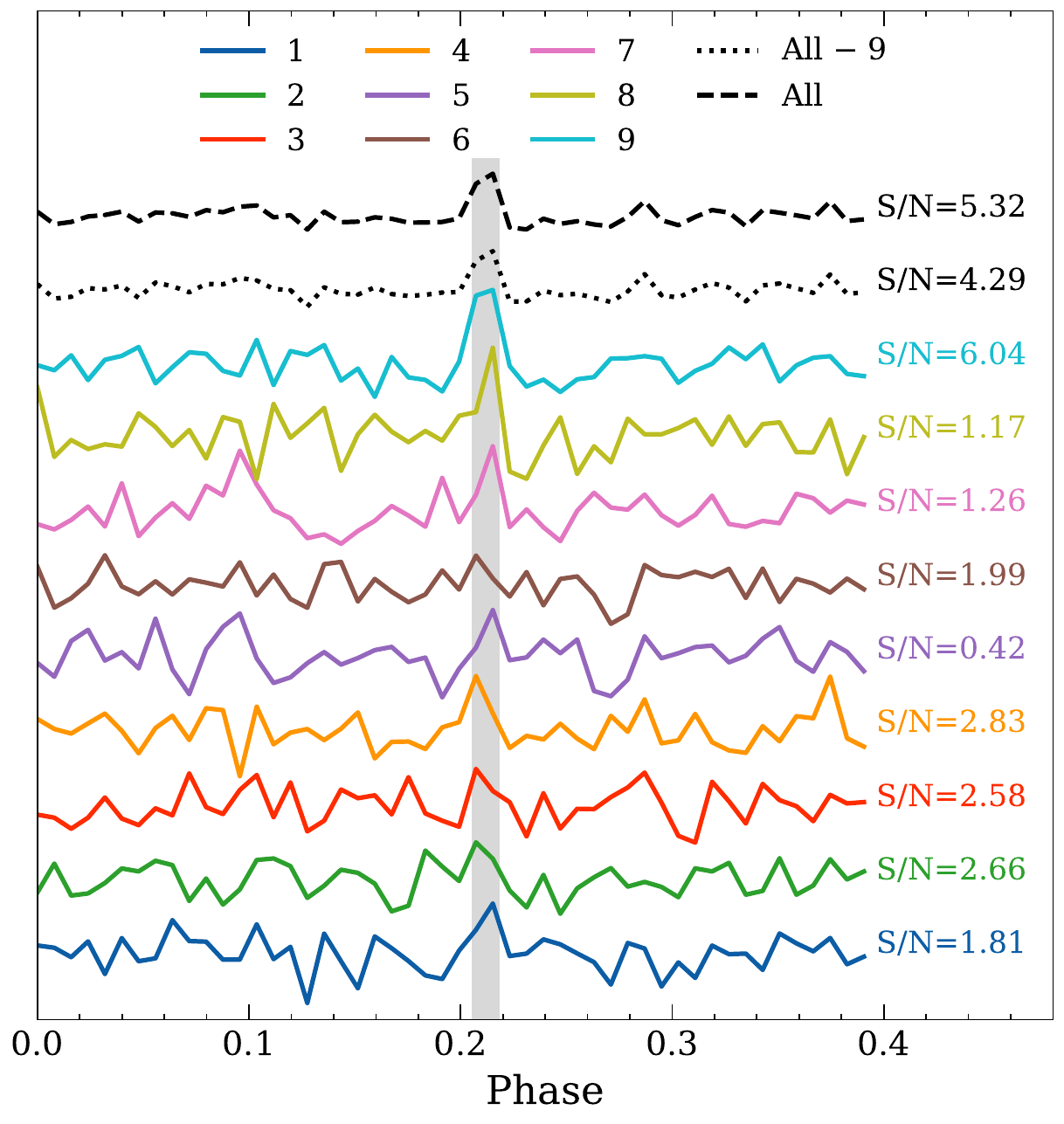}
  \caption{Nine pulse profiles of 192-s subintegrations after removing the single pulses with $\mathrm{S/N}\ge5$. Subintegration 9 has a higher $\mathrm{S/N}$ (6.04) than the other subintegrations. If we remove subintegration 9 and add up the remaining eight subintegrations, we still detect a pulse in the integrated pulse profile with $\mathrm{S/N}=4.29$. The pulse profiles of each subintegration and the total integration are plotted width different line styles, as denoted in the figure. The grey shaded area is the on-pulse region similar to \FIG{fig:inte_profile}, and the integration time per data point is 9.7 ms, which is eight times of sampling time in the bottom panel of \FIG{fig:inte_profile}. %the integration time per data point is the same as \FIG{fig:Intensity-Phsae-Time}.
  \label{fig:running}}
\end{figure}

Reviewing the $\mathrm{S/N}$ distribution in \FIG{fig:SN-fitting} and the peak flux density distribution in \FIG{fig:flux-fitting}, both the $\mathrm{S/N}$ distribution and the peak flux distribution are described by a mixture of subpopulations. It is obvious that the subpopulation with $\mathrm{S/N}\sim 1$ or flux close to the detection threshold is due to the radiometer noise, and the other subpopulations represent the single-pulse signals from \NAME J0628+0909. The single pulse population overlaps with the population of the radiometer noise. Thus it is possible that some weak single pulses were buried in the noise and were not picked up in the single-pulse search process. To examine the weak pulses, we remove any single pulse with $\mathrm{S/N}\ge5$ and form the 192-s subintegration shown in \FIG{fig:running}. The $\mathrm{S/N}$s for all the subintegrations are rather low ($\mathrm{S/N}\sim 2$) except for the last subintegration, which has $\mathrm{S/N}=6.04$. Those low-$\mathrm{S/N}$ subintegrations, i.e. the subintegrations 1 to 8, produce a total $\mathrm{S/N}=4.29$ after time integration, showing that there are low-amplitude single pulses buried below the radiometer noise floor. These weak single pulses seem to be distributed uniformly in time, as the subintegrations 1 to 8 all had a similar level of $\mathrm{S/N}$. For subintegration 9, $\mathrm{S/N}=6.04$, which is stronger by a factor of 2 -- 4 compared to the subintegration 1 to 8. This may be caused by the scintillation amplification, because the scintillation time is comparable to the subintegration length (also see the discussion in Section~\ref{sec:sigps}).

\subsection{waiting-time distribution}\label{sec:wait time}
\begin{figure}
\centering
\includegraphics[width=\hsize]{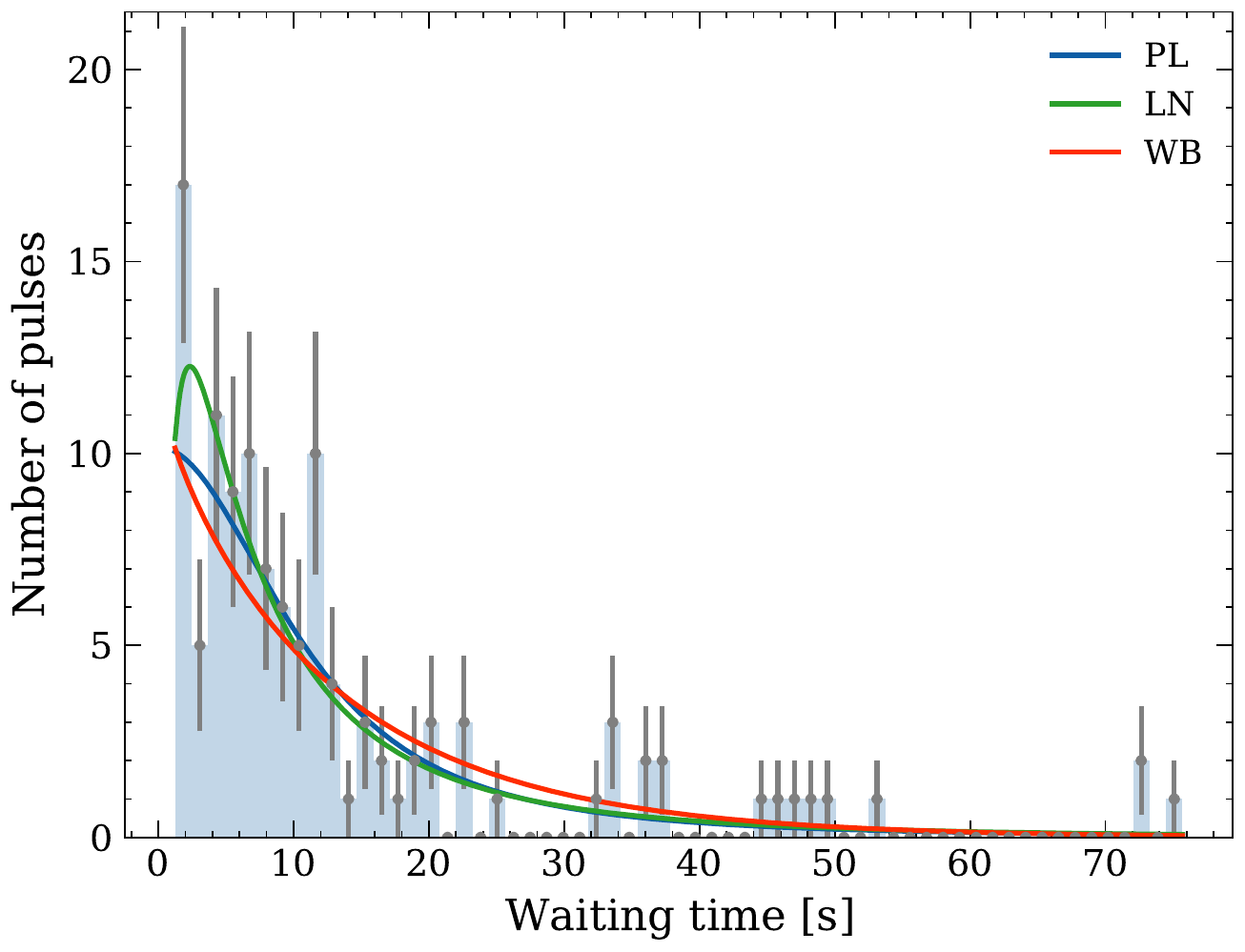}
  \caption{Measured waiting-time distribution and the modellings. Our models with power law (PL, blue),  log-normal (LN, green), and Weibull (WB, red) distribution are also shown. The error bars are Poissonian, i.e. $\sqrt{N}$.
  \label{fig:wait}}
\end{figure}

We measured the waiting-time between two successive single pulses with $\mathrm{S/N}\ge7$. The distribution is shown in \FIG{fig:wait}, where our modellings with PL, LN, and Weibull (WB) distributions are compared. Here, the WB probability density distribution function is
\citep{Oppermann_2018}
\begin{equation}
    \displaystyle {f_{\rm WB}\left(x\mid\lambda, k\right)= \begin{cases}\lambda{k}\left(\lambda x\right)^{k-1}\exp\left[-\left(\lambda{x}\right)^{k}\right] & x\ge0, \\ 0 & x<0,\end{cases}}
\end{equation}
where $k$ is the shape parameter, and $\lambda>0$ is the expected event rate.

\TAB{tab:Wait-fitting} summarises the inferred model parameters. We note that none of the models can account for the extended tail in the waiting-time distribution. However, according to the KS test $p$-values, all the models are still acceptable given the limited number of single pulses. We estimate the event rate from the WB distribution, which is $270\substack{+35 \\-29}~\rm{h^{-1}}$. The shape parameter of the WB distribution is very close to 1, which shows that the occurrence of single pulses follows the Poissonian process; that is, there is no temporal correlation between the single pulses.
\begin{table}
\centering
\caption{\label{tab:Wait-fitting}
Inferred model parameters for the distribution of single-pulse waiting-times.}
\begin{tabular}{cccc}
\hline\hline \\
\multicolumn{4}{c}{PL ($\rm M_1$)} \\ [1.5ex]
$x_0$ [s] & $\alpha$ & $p$-value & $\log_{10}{\cal B}$ \\ [1.5ex]
\hline \\
$13.55\substack{+1.45\\-3.29}$  & $2.89\substack{+0.11\\-0.46}$ & $0.31$  & $-0.23$ ($\rm M_1$/$\rm M_2$) \\ [1.5ex]
\hline\hline \\
\multicolumn{4}{c}{LN ($\rm M_2$)} \\ [1.5ex]
$\mu$ [s] & $\sigma$ [s] & $p$-value & $\log_{10}{\cal B}$ \\ [1.5ex]
\hline \\
$2.04\substack{+0.15\\-0.15}$  & $1.09\substack{+0.12\\-0.10}$ & $0.20$   & $0$ ($\rm M_2$/$\rm M_2$) \\ [1.5ex]
\hline\hline \\
\multicolumn{4}{c}{WB ($\rm M_3$)} \\ [1.5ex]
$\lambda$ [1/s] & $k$ & $p$-value & $\log_{10}{\cal B}$ \\ [1.5ex]
\hline \\
$0.08\substack{+0.01\\-0.01}$ & $0.97\substack{+0.08\\-0.02}$ & $0.09$   & $-1.35$ ($\rm M_3$/$\rm M_2$)  \\ [1.5ex]
\hline\hline
\end{tabular}
\end{table}

\begin{figure}
\centering
\includegraphics[width=\hsize]{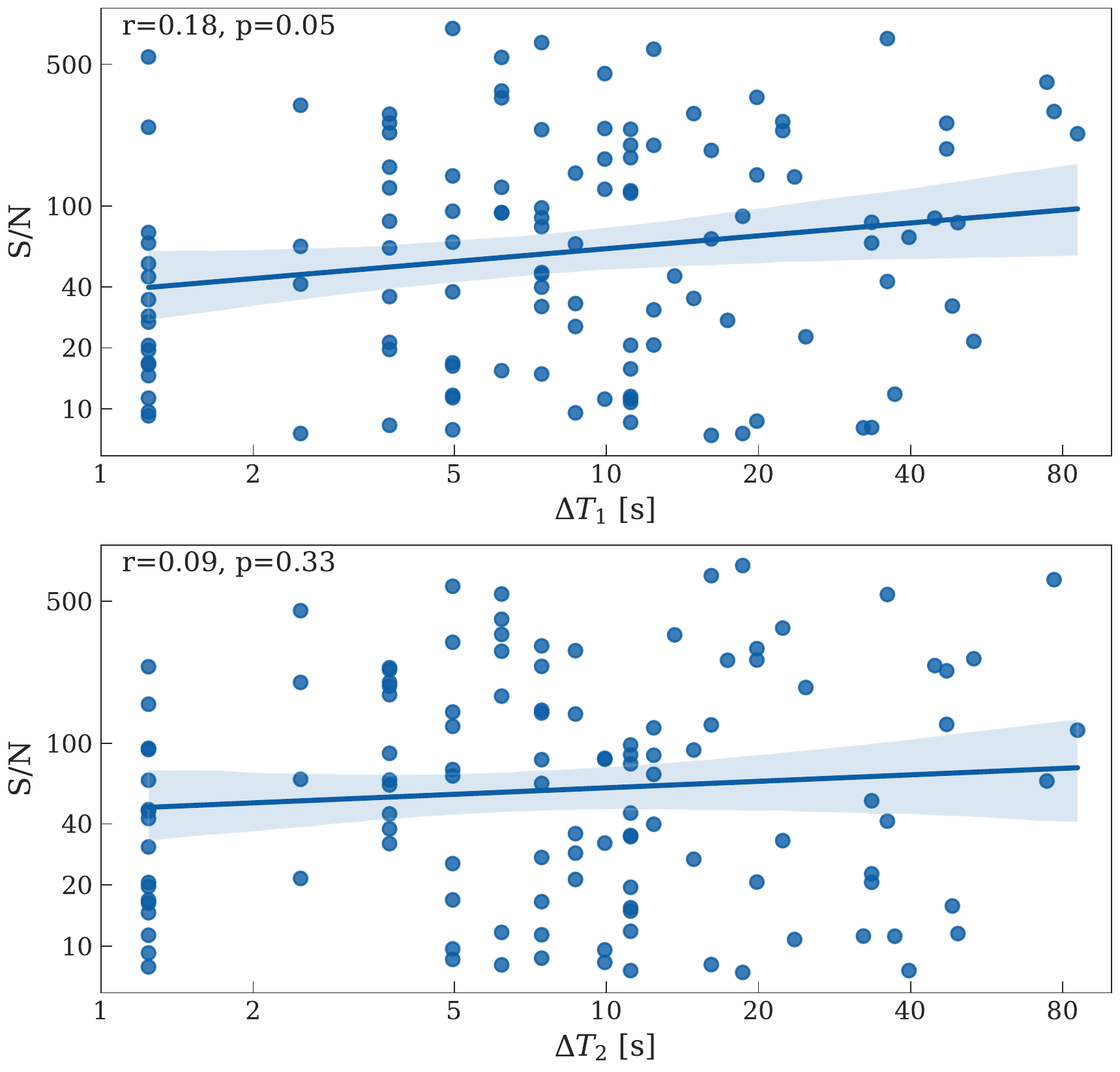}
  \caption{Every single pulse is plotted with a solid dot with a bivariate distribution of waiting-times and $\mathrm{S/N}$. The top panel is for the waiting-time $\Delta T_{1}$, the time from the previous pulse to the given pulse we compute energy for; the lower panel is for the waiting-time $\Delta T_{2}$, the time from the given pulse to the next pulse. The blue solid line is the linear regression, and the shaded region represents the 95\% confidence interval. The correlation coefficient and corresponding $p$-value are given in the plots. No significant correlation is detected in either case.
  \label{fig:correlation}}
\end{figure}

We tried to measure the correlation between the energy of single pulses and waiting-time. Two correlations were measured, (1) the correlation between the pulse energy and waiting-time from the previous pulse to the given pulse ($\Delta T_{1}$) and (2) the correlation between the pulse energy and waiting-time from the given pulse to the next pulse ($\Delta T_{2}$).  As shown in \FIG{fig:correlation}, the correlations are weak, with Pearson's coefficient being $r=0.18$ for $\Delta T_{1}$ and $r=0.09$ for $\Delta T_{2}$.

\section{Discussion and conclusions}\label{sec:discuss}
In this work, we analysed the half-hour FAST observation of \NAME J0628+0909 with a central frequency of 1250 MHz and a bandwidth of 500 MHz. We conducted single-pulse studies and measured the polarisation properties of the source. The peak flux distribution of single pulses was measured, and we concluded that three log-normal components are required to describe the distribution. We noted that there are low-flux pulses buried under the noise floor, and it seems that they are continuous in time. We found that the correlation between the waiting-time and the pulse energy is relatively weak.

Prior studies have noted that the pulse energies of the majority of RRATs follow a log-normal distribution, with a few showing power law tails \citep[e.g.][]{Mickaliger_2018, Shapiro-Albert_2018, Meyers_2019}. \cite{Mickaliger_2018} discovered that 12 of the 14 RRATs they analysed exhibited “bumps” in their energy distributions, which can be fitted with two distinct log-normal distributions. However, \citet{Shapiro-Albert_2018} looked at the same RRATs but did not discover any “bumps”. According to their explanation, analysing data with multiple epochs may bridge the gap between two distribution populations. In this paper, the $\mathrm{S/N}$ and peak flux distributions for \NAME J0628+0909 were computed. We note that multiple components are required. In our case of \NAME J0628+0909, we found that the mixture model of a lower-$\mathrm{S/N}$ distribution (the Gaussian normal) and two log-normal distributions is favoured. The lower-energy population is identified as a noise contribution, and the two log-normal populations are from the RRAT emission. The Bayes factor ratio indicates that the mixture of two log-normal populations is preferred over one population ($\log_{10}{\cal B}=1.63$), although the KS test could not discriminate between the two models. In this case, the probability models are nested, and the Bayes factor is well defined. It is not surprising that Bayes analysis is more sensitive than the non-parametric statistics, namely the KS test. 

Similarly, we note that multiple components are required to fit the distribution of peak flux; that is, the model with three log-normal components is significantly preferred to that with two log-normal components (Bayes factor $\log_{\rm 10} {\cal B}=12.51$). Previous research found that the peak flux of RRAT pulses may follow a power law distribution or a log-normal distribution \citep[e.g.][]{McLaughlin_2006, Brylyakova_2021, Tyulbashev_2021}, although \cite{CBM17} compared a power law model with a log-normal model and found that the latter had a better fit. Our results do not conflict with those of previous studies. The FAST 3-$\sigma$ sensitivity for the sampling time of 49~$\mathrm{\mu s}$ and bandwidth of 500 MHz is 16 mJy, which is approximately the central value of the distribution components we detected as shown in Figure~\ref{fig:SN-fitting} and \ref{fig:flux-fitting}. The components we detected are at least a few times weaker than the previous results. Our results hint that the intrinsic pulse peak flux or energy distributions are more complex if observed with higher sensitivity. We hypothesise that the requirement for the mixture of multiple populations is merely the consequence of approximating the more complex intrinsic distribution.

We noticed weaker pulse signals (below the 5-$\sigma$ threshold of 28 mJy) buried under the radiometer noise floor. An integrated pulse profile can be found once we fold the data.  Surveys such as that conducted by \cite{CBM17} and \cite{Tyulbashev_2021} reported similar weak pulses in other RRATs, and all of these provide evidence to support the view that RRATs are low-flux pulsars \citep{Weltevrede_2006}. Furthermore, our detection of the weaker pulse population and the related integrated pulse profile show that the RRAT does \emph{not} stop emission in the radio between the strong pulses with $\mathrm{S/N}\ge5$. The average flux of the weaker pulse is, indeed, much weaker than the strong pulses. In our sample, the 30-min integration of weaker pulses produces $\mathrm{S/N}=5.32$ (see \FIG{fig:running}), and the average flux will be approximately 3000 times weaker than the bright burst we detected (single pulse $\mathrm{S/N}\simeq 800$). However, as our detected single-pulse distribution also extends to the low-$\mathrm{S/N}$ regime, we cannot conclude, at this stage, if the weaker pulses form another independent population or the weaker-pulse population can be separated from the single-pulse population. 

The event rate for single pulses with $\mathrm{S/N}\ge7$ is $270\substack{+35 \\-29}~\mathrm{h^{-1}}$. The burst rate of \NAME J0628+0909 was previously measured at the Arecibo Observatory \citep{Deneva_2009}, where 42 single pulses with $\mathrm{S/N}\ge5$ were detected in the 1072-s observation. The corresponding event rate was 141 $\mathrm{h^{-1}}$. The difference is probably caused by the selection bias induced by the telescope sensitivity. The Arecibo Observatory parameters are: an effective gain of $G\approx 10.4\,\mathrm{K\, Jy^{-1}}$, a typical system noise temperature of $T_{\rm sys}\approx30$ K, a bandwidth of $\Delta\nu=100$ MHz with a central frequency of $\nu=1440$ MHz, and $n_{\rm pol}=2$. Correcting the flux density difference due to the central frequency offset with $S_\nu\propto\nu^{-0.75}$ \citep{Nice_2013}, and recomputing $\mathrm{S/N}$ by fixing the boxcar width $w$ as 10 ms as adopted in \citet{Deneva_2009}, the single pulses with $\mathrm{S/N}\ge5$ of \cite{Deneva_2009} approximately correspond to $\mathrm{S/N}\ge20$ at FAST. If we raise the detection threshold to $\mathrm{S/N}\ge20$, the burst rate will be 172 $\mathrm{h^{-1}}$, roughly compatible with the Arecibo results considering the Poissonian error.

We note that among the PL, LN and WB distributions, LN describes the waiting-time best according to the Bayes factor, similar to the case of soft gamma-ray repeaters in the high-energy band \citep{GEW99}. However, unlike in the case of soft gamma-ray repeaters, we find little correlation between pulse waiting-time and pulse energy, as shown in \FIG{fig:correlation}. The RRATs may have different mechanisms to produce the strong single pulses rather than via the energy store-release scenario. Similar analyses on three RRATs \citep{Shapiro-Albert_2018} and one on nulling pulsars \citep{Gajjar_2012} have yielded consistent conclusions.

%--------------------------------------------------------------------

\section*{Acknowledgements}
This work made use of the data from FAST (Five-hundred-meter Aperture Spherical radio Telescope). FAST is a Chinese national mega-science facility, operated by National Astronomical Observatories, Chinese Academy of Sciences. This work is supported by the National SKA Program of China (2020SKA0120100), the National Key R\&D Program of China (2017YFA0402602), the National Nature Science Foundation grant no. 12041303, the CAS-MPG LEGACY project, and funding from the Max-Planck Partner Group.

%%%%%%%%%%%%%%%%%%%%%%%%%%%%%%%%%%%%%%%%%%%%%%%%%%
\section*{Data Availability}
% The inclusion of a Data Availability Statement is a requirement for articles published in MNRAS. Data Availability Statements provide a standardised format for readers to understand the availability of data underlying the research results described in the article. The statement may refer to original data generated in the course of the study or to third-party data analysed in the article. The statement should describe and provide means of access, where possible, by linking to the data or providing the required accession numbers for the relevant databases or DOIs.

The original observation data will be available from the FAST Data Center (see \url{https://fast.bao.ac.cn/cms/article/129/}). The default data proprietary period is 18 months. Detailed access information can be found on the website of the FAST Data Center. Furthermore, the processed data for plotting can be obtained from our website (see \url{https://psr.pku.edu.cn/upload/xra/published_data/}).



\bibliographystyle{mnras}
\bibliography{ref} % if your bibtex file is called example.bib

%\begin{thebibliography}{99}
%\end{thebibliography}

%%%%%%%%%%%%%%%%%%%%%%%%%%%%%%%%%%%%%%%%%%%%%%%%%%

%%%%%%%%%%%%%%%%% APPENDICES %%%%%%%%%%%%%%%%%%%%%

% \appendix

% \section{Some extra material}

% If you want to present additional material which would interrupt the flow of the main paper,
% it can be placed in an Appendix which appears after the list of references.

%%%%%%%%%%%%%%%%%%%%%%%%%%%%%%%%%%%%%%%%%%%%%%%%%%

% Don't change these lines
\bsp	% typesetting comment
\label{lastpage}
\end{document}